_____________________________________________________________________
\documentstyle[12pt]{article}
\begin{document}

\title{Exact Solutions of the two-dimensional Schr\"{o}dinger \\
equation with certain central potentials}

\author{Shi-Hai Dong\thanks{Electronic address: dongsh@phys. ksu. edu}\\
{\footnotesize Physical and Theoretical Chemistry Laboratory, University
of
Oxford, Oxford OX1 3QZ, UK}\\
{\footnotesize  and Department of Physics, Cardwell Hall, 
Kansas State University, Manhattan, Kansas 66506}}

\date{}
\maketitle

\begin{abstract}

By applying an ansatz to the eigenfunction, 
an exact closed form solution of the Schr\"{o}dinger 
equation in 2D is obtained with the potentials, 
$V(r)=ar^2+br^4+cr^6$, $V(r)=ar+br^2+cr^{-1}$ and 
$V(r)=ar^2+br^{-2}+cr^{-4}+dr^{-6}$, respectively. 
The restrictions on the parameters of the given potential and the 
angular momentum $m$ are obtained.


\end{abstract}

\vskip 1cm

\newpage

\begin{center}
{\large 1. Introduction}
\end {center}
 
One of the important tasks of quantum mechanics is to solve 
the Schr\"{o}dinger equation with the physical potentials. 
It is well known that the exact solution
of the Schr\"{o}dinger equation
are possible only for the certain potentials such as Coulomb, harmonic
oscillator potentials. Some approximation methods
are frequently used to obtain the solution. 
In the past several decades, many efforts have been produced in the literature
to study the stationary Schr\"{o}dinger equation with the  
central potentials containing negative powers of the 
radial coordinate [1-31]. Generally, 
most of authors carried out these problems in the 
three-dimensional space. Recently, the study of higher 
order central potentials has 
been much more desirable 
to physicists and mathematicians, who want to understand 
a few newly discovered physical phenomena 
such as structural phase transitions [1], polaron
formation in solids [2] and the 
concept of false vacuo in filed 
theory [3]. Besides, the solution of the Schr\"{o}dinger equation 
with the sextic potential $V(r)=ar^2+br^4+cr^6$ can be applied in the 
field of fibre optics [4], where one wants to solve a similar problem of an 
inhomogeneous spherical or circular wave guide with refractive index 
profile function of the sextic-type potential. Its solution is also
applicable to molecular physics [5]. 
The study for the mixed potential $V(r)=a_{1}r+b_{1}r^2+c_{1}r^{-1}$ 
(harmonic+linear+Coulomb) as a phenomenological potential 
can be used in nuclear physics. However, the study
for the singular even-power potential 
$V(r)=ar^2+br^{-2}+cr^{-4}+dr^{-6}$ has been widely 
used in the different fields such as atomic physics and
optical physics [29-31]. 
Actually, interest in these 
anharmonic oscillator-like interactions stems from the fact 
that the study of the relevant Schr\"{o}dinger equation, for 
example, in the atomic and molecular physics as well as 
nuclear physics , provides us with 
insight into the physical problem in question. 

With the wide interest in the lower-dimensional 
field theory in the recent literature, however, it is necessary 
to study the two-dimensional Schr\"{o}dinger equation with 
the certain central potentials such as the  
sextic and mixed potentials as well as the 
singular even-power potential, an
investigation which, to our knowledge, 
has never been appeared in the literature. 
Furthermore, two-dimensional models are often applied to
make the more involved higher-dimensional systems tractable. 
Therefore, it seems reasonable to study the two-dimensional
Schr\"{o}dinger equation with these potentials, 
which is the purpose of this paper. On the other hand, we have 
succeeded in studying the two-dimensional Schr\"{o}dinger 
equation with some anharmonic potentials [16, 17]. 

This paper is organized as follows. 
Section 2 studies the solution 
of the two-dimensional Schr\"{o}dinger 
equation with the sextic potential $V(r)=ar^2+br^4+cr^6$
using~~an~~ansatz~~for~~the~~eigenfunction. 
The~~study~~ for~~ the~~ mixed~~ potential $V(r)=a_{1}r+b_{1}r^2+c_{1}r^{-1}$
will be presented in section 3. In section 4, we will study the  
singular even-power potential 
$V(r)=ar^2+br^{-2}+cr^{-4}+dr^{-6}$. 
A brief conclusion will be given in the last section 5. 

\vskip 1cm
\begin{center}
{\large 2. The Study for the Sextic Potential}
\end{center}

Throughout this paper the natural units $\hbar=1$ 
and $\mu=1/2$ are employed. Consider the two-dimensional 
Schr\"{o}dinger equation with a potential $V(r)$
that depends only on the distance $r$ from the origin
$$H\psi =-\left(
\displaystyle {1 \over r} \displaystyle {\partial \over \partial r} 
r \displaystyle {\partial \over \partial r} + 
\displaystyle {1 \over r^{2}} \displaystyle {\partial^{2} \over 
\partial \varphi^{2} } \right)\psi +V(r) \psi =E \psi, \eqno(1)$$ 

\noindent
where the potential is taken as
$$V(r)=ar^2+br^4+cr^6. \eqno(2)$$

\noindent
The choice of $r, \varphi$ coordinates reflects a model where the
full Hilbert space is the tensor product of the space of 
square-integrable functions on the positive half-line with the
space of square-integrable functions on the circle. We
therefore write
$$\psi({\bf r}, \varphi)=r^{-1/2} R_{m}(r) e^{ \pm im \varphi}, 
~~~~~m=0, 1, 2, \ldots \eqno (3) $$  

\noindent
and this factorization leads to a second-order equation for the
radial function $R_{m}(r)$ with vanishing coefficient of the 
first derivative, i. e. 
$$\displaystyle {d^{2} R_{m}(r) \over dr^{2} }
+\left[E-V(r)-\displaystyle {m^{2}-1/4 \over r^{2}} \right] R_{m}(r)
=0, \eqno (4) $$

\noindent
where $m$ and $E$ denote the 
angular momentum and energy, respectively. 
For the solution of Eq. (4), we make an ansatz [6-21]
for the radial wave function 
$$R_{m}(r)=\exp[p_{m}(r)]\sum\limits_{n=0}a_{n}r^{2n+\delta}, \eqno(5)$$
 
\noindent
where
$$p_{m}(r)=\frac{1}{2} \alpha r^2
+\frac{1}{4}\beta r^4. \eqno(6)$$

\noindent
Substituting Eq. (5) into Eq. (4) and equating the coefficient of 
$r^{2n+\delta+2}$ to zero, one can obtain 
$$A_{n}a_{n}+B_{n+1}a_{n+1}+C_{n+2}a_{n+2}=0 \eqno(7)$$

\noindent
where 
$$A_{n}=\alpha^{2}+(3+2\delta+4n)\beta-a \eqno(8a)$$
$$B_{n}=E+(1+2\delta+4n)\alpha \eqno(8b)$$
$$C_{n}=(\delta+2n)(-1+\delta+2n)-(m^2-1/4) \eqno(8c)$$

\noindent
and
$$\beta^{2}=c \eqno(9a)$$
$$2\alpha\beta=b. \eqno(9b)$$

\noindent
It is easy to obtain the 
values of parameters for $p_{m}(r)$
from the Eq. (9) written as
$$\beta=\pm \sqrt{c}, ~~~\alpha=\frac{b}{2\beta}. \eqno(10)$$

If the first non-vanishing coefficient $a_{0}\not=0$ in Eq. (7), and so 
we can obtain $C_{0}=0$ from Eq. (8c), i. e. $\delta=-m+1/2$ or $m+1/2$. 
In order to retain the well-behaved solution 
at the origin and at infinity, we choose 
$\delta$ and $\beta$ as follows:
$$\delta=m+1/2, ~~~~\beta=-\sqrt{c}, \eqno(11a)$$ 

\noindent
from which, one can obtain 
$$\alpha=-\frac{b}{2\sqrt{c}}, \eqno(11b)$$ 

\noindent
On the other hand, if the $p$th non-vanishing coefficient $a_{p}\not=0$, but
$a_{p+1}=a_{p+2}=a_{p+3}=\cdots=0$, 
it is easy to obtain $A_{p}=0$ from Eq. (8a), i. e. 
$$a+2\sqrt{c}(2+m+2p)-\frac{b^2}{4c}=0, \eqno(12)$$
which is a restriction on the parameters $a, b, c$ of the 
potential and angular momentum $m$ and $p$ ($p\leq n$). As we 
know, $A_{n}, B_{n}$ and $C_{n}$ must 
satisfy the determinant relation for a nontrivial solution
$$\det\left|
\begin{array}{llllll}
B_{0} & C_{1}&\cdots&\cdots&\cdots&0\\
A_{0}& B_{1}& C_{2}& \cdots &\cdots&0\\
\vdots&\vdots&\vdots&\ddots&\vdots&\vdots\\
0&0&0&0&A_{p-1}&B_{p}\\
\end{array}\right|=0. \eqno(13)$$
 
In order to expound this method, we will give the 
exact solutions for the different $p=0, 1$ as follows. 

(1): when $p=0$, it is easy to obtain $B_{0}=0$ from Eq. (13), 
which, together with Eq. (11), leads to 
$$E_{0}=\frac{b(1+m)}{\sqrt{c}}. \eqno(14)$$

\noindent
In this case, however, the restriction on 
the parameters of the potential and the angular momentum $m$
will be obtained as 
$$a+2\sqrt{c}(2+m)-\frac{b^2}{4c}=0. \eqno(15)$$

\noindent
The corresponding eigenfunction for $p=0$ can now be read as
$$R_{m}^{(0)}=a_{0}r^{\delta}
\exp\left[\displaystyle{-\frac{b}{4\sqrt{c}}}r^2-
\displaystyle{\frac{\sqrt{c}}{4}}r^4\right], \eqno(16)$$

\noindent
where $a_{0}$ is the normalization constant and $\delta$ is given by Eq. (11). 



(2): when $p=1$, one can arrive at the following relation from Eq. (13), 
$$B_{0}B_{1}-A_{0}C_{1}=0 \eqno(17)$$
from which  we can obtain
$$E_{1}=\frac{b(2+m)}{\sqrt{c}}\pm 
\frac{\sqrt{b^2(2+m)-4c(1+m)(2+2\sqrt{c}(2+m))}}
{\sqrt{c}}.\eqno(18)$$

\noindent
However, the corresponding restriction on the parameters and $m$ 
can be obtained as
$$a+2(4+m)\sqrt{c}-\frac{b^2}{4c}=0. \eqno(19)$$. 
The corresponding eigenfunction for $p=1$ can be read as
$$R_{m}^{(1)}=(a_{0}+a_{1}r^2)r^{\delta}
\exp\left(\displaystyle{-\frac{b}{4\sqrt{c}}}r^2-
\displaystyle{\frac{\sqrt{c}}{4}}r^4\right), \eqno(20)$$

\noindent
where $\delta$ has been given by Eq. (11), the coefficients 
$a_{0}$ and $a_{1}$ can be determined by the normalization condition 
completely. 

Following this way, we can generate a class of exact solutions 
through setting $p=1, 2,\cdots, $etc.  For the general case, 
if the $p$th non-vanishing coefficient $a_{p}\not=0$, but
$a_{p+1}=a_{p+2}=\cdots=0$, from which we can obtain $A_{p}=0$, i. e. 
$$\alpha^2+(3+2\delta+4p)=a. \eqno(21)$$

\noindent
The corresponding eigenfunction can be written as
$$R_{m}^{(p)}=(a_{0}+a_{1}r^2+\cdots+a_{p}r^{2p})r^{\delta}
\exp\left[\displaystyle{-\frac{b}{4\sqrt{c}}}r^2-
\displaystyle{\frac{\sqrt{c}}{4}}r^4\right], \eqno(22)$$

\noindent
where $\delta$ has been given by Eq. (11a), 
and $a_{i} (i=1, 2, \cdots p)$, can be expressed by 
recurrence relation Eq. (7) and in principle obtained by the normalization 
condition. 

\vskip 1cm
\begin{center}
{\large 3. The Study for the Mixed Potential}
\end{center}

The study for this potential is similar to that 
for sextic potential except for
taking the ansatz as 
$$R_{m}(r)=\exp[p_{m}(r)]\sum\limits_{n=0}a_{n}r^{n+\delta}, \eqno(23)$$

\noindent
where $p_{m}$ is taken as
$$p_{m}(r)=\alpha r+\frac{1}{2}\beta r^2. \eqno(24)$$

\noindent
We can solve the two-dimensional 
Schr\"{o}dinger equation with the this potential 
$$V(r)=ar+br^{2}+\frac{c}{r}. \eqno(25)$$
Similarly, we can obtain the following sets of equations 
after substituting Eq. (23) into Eq. (4) and equating the coefficients of 
$r^{\delta+n}$ to zero, 
$$A_{n}a_{n}+B_{n+1}a_{n+1}+C_{n+2}a_{n+2}=0 \eqno(26)$$
where
$$A_{n}=E+\beta(1+2n+2\delta) \eqno(27a)$$
$$B_{n}=-c+\alpha(2n+2\delta) \eqno(27b)$$
$$C_{n}=(n+\delta)(-1+n+\delta)-(m^2-1/4) \eqno(27c)$$
and
$$\beta^{2}=b, ~~~~~~~~~2\alpha\beta=a. \eqno(27d)$$
Similar to the above choices, we can choose $\beta$ and $\delta$ as $-\sqrt{b}$
and $m+1/2$, respectively. According to these choices, the parameter 
$\alpha$ can be obtained as
$$\alpha=-\frac{a}{2\sqrt{b}}. \eqno(28)$$

Now, let us consider the case $a_{p}\not=0$, but $a_{p+1}=a_{p+2}=\cdots=0$, 
then we can get $A_{p}=0$. In this case, the energy eigenvalue can be written
as 
$$E_{p}=2\sqrt{b}(1+m+p). \eqno(29)$$

\noindent
Likewise, the nontrivial solution of recursion relation Eq. (26) can be 
obtained by Eq. (13). 
The exact solutions for $p=0$ and $p=1$ can be discussed below. 

(1): when $p=0$, we can arrive at
$$E_{0}=2\sqrt{b}(1+m) \eqno(30)$$. 

\noindent
and $B_{0}=0$, i. e. 
$$2c\sqrt{b}=a(1+2m), \eqno(31)$$
which is a restriction on the corresponding parameters of the potential 
and the angular momentum $m$. 
The eigenfunction, however, can be read as
$$R_{m}^{(0)}=a_{0}r^{\delta}\exp\left[-\frac{ar+br^2}{2\sqrt{b}}\right], 
\eqno(32)$$
where $\delta$ is taken as $m+1/2$, 
the coefficient $a_{0}$ can be evaluated by the normalization condition. 

(2): when $p=1$, the energy eigenvalue can be written as
$$E_{1}=2\sqrt{b}(2+m). \eqno(33)$$
Moreover, we can obtain the restriction on the parameters of the potential 
and the angular momentum $m$ from the determinant relation Eq. (13) as 
$B_{0}B_{1}=A_{0}C_{1}$, i. e. 
$$\left\{c+\frac{(1+2m)a}{2\sqrt{b}}\right\}
\left\{c+\frac{(3+2m)a}{2\sqrt{b}}\right\}
=2\sqrt{b}(1+2m). \eqno(34)$$
In this case, the corresponding eigenfunction can be written as
$$R_{m}^{(1)}=(a_{0}+a_{1}r)r^{\delta}
\exp\left[-\frac{ar+br^2}{2\sqrt{b}}\right], \eqno(35)$$
where $a_{0}$ and $a_{1}$ can be obtained by the recursion relation Eq. (26) 
and the normalization relation. 

Similarly, if $a_{p}\not=0$, but $a_{p+1}=a_{p+2}=\cdots=0$, 
we can get $A_{p}=0$. In this case, the eigenfunction can be written as
$$R_{m}^{(p)}=(a_{0}+a_{1}r+\cdots+a_{p}r^{p})r^{\delta}
\exp\left[-\frac{ar+br^2}{2\sqrt{b}}\right], \eqno(36)$$

\noindent
where $\delta$ is taken as $m+1/2$, and 
the coefficients $a_{i} (i=1, 2, \cdots, p)$ 
can be calculated by Eq. (26) and the normalization condition. 

\vskip 1cm
\begin{center}
{\large 4. The Study for the Singular Even-Power Potential}
\end{center}

Similar to above discussion, 
the study for the central singular even-power potential
can be taken the following ansatz  
$$R_{m}(r)=\exp[p_{m}(r)]\sum\limits_{n=0}a_{n}r^{2n+\delta}, \eqno(37)$$
where $p_{m}$ is taken as
$$p_{m}(r)=\frac{1}{2}\alpha r^2+\frac{1}{2}\beta r^{-2}. \eqno(38)$$

\noindent
We can solve the two-dimensional 
Schr\"{o}dinger equation with the potential 
$$V(r)=ar^2+\frac{b}{r^2}+\frac{c}{r^4}+\frac{d}{r^6}. \eqno(39)$$
Likewise, one can get the following sets of equations after substituting the 
ansatz Eq. (37) into Eq. (4) and equating the coefficients of 
$r^{\delta+n}$ to zero, 
$$A_{n}a_{n}+B_{n+1}a_{n+1}+C_{n+2}a_{n+2}=0, \eqno(40)$$
where
$$A_{n}=E+\alpha(1+2\delta+4n) \eqno(41a)$$
$$B_{n}=-b-2\alpha\beta-(m^2-1/4)+(\delta+2n)(-1+\delta+2n) \eqno(41b)$$
$$C_{n}=(3-2\delta-4n)-c \eqno(41c)$$
and
$$\alpha^{2}=a, ~~~~~~~~~\beta^2=d. \eqno(42)$$

\noindent
Similar to the above choices, we can choose $\alpha$ and $\beta$ as $-\sqrt{a}$
and $-\sqrt{d}$, respectively. 

\noindent
Moreover, if $a_{0}\not=0$, then one can obtain $C_{0}=0$, i. e. 
$$\delta=(3/2+\mu), \eqno(43)$$
where $\mu\equiv\frac{c}{2\sqrt{d}}$. 
However, if $a_{p}\not=0$, but $a_{p+1}=a_{p+2}=\cdots=0$, 
then it leads to $A_{p}=0$, from which one can obtain the energy eigenvalue
as 
$$E_{p}=\sqrt{a}(4+4p+2\mu). \eqno(44)$$

\noindent
In the following section, let us discuss the  corresponding exact solutions 
for $p=0$ and $p=1$. 

(1): when $p=0$, we can arrive at
$$E_{0}=\sqrt{a}(4+2\mu). \eqno(45)$$
In this case, it means that $B_{0}=0$ from the determinant relation Eq. (13), 
which will lead to the constraint condition between the parameters of the 
potential and the angular momentum quantum number $m$, 
$$(1+\mu)^{2}-b-2\sqrt{ad}-m^2=0. \eqno(46)$$
The corresponding eigenfunction, however, can be written as
$$R_{m}^{(0)}=a_{0}r^{\delta}\exp\left[-\frac{\sqrt{a}r^2+\sqrt{d}r^{-2}}{2}
\right], \eqno(47)$$
where and hereafter $\delta$ is given by Eq. (43) and the coefficient 
$a_{0}$ can be obtained by the normalization condition. 

(2): $p=1$, the energy eigenvalue can be obtained from Eq. (44) as 
follows
$$E_{1}=\sqrt{a}(8+2\mu). \eqno(48)$$
In this case, the determinant relation Eq. (13) means that 
$B_{0}B_{1}=A_{0}C_{1}$, which will result in the following 
restriction on the parameters and angular momentum quantum $m$, 
$$\left[-b-2\sqrt{ad}+(1+\mu)^2-m^2\right]
\left[-b-2\sqrt{ad}+(3+\mu)^2-m^2\right]-16\sqrt{ad}=0. \eqno(49)$$
The eigenfunction for $p=1$ can be read as
$$R_{m}^{(1)}=(a_{0}+a_{1}r^2)r^{\delta}
\exp\left[-\frac{\sqrt{a}r^2+\sqrt{d}r^{-2}}{2}\right], \eqno(50)$$
where and hereafter $\delta$ is given by Eq. (43), 
and $a_{i} (i=0, 1)$ can be calculated from Eq. (40) and 
the normalization relation. 
Following this method, we can obtain a class of exact solutions through setting
the different $p$. Generally, the corresponding eigenfunction for $p$ can be 
written as 
$$R_{m}^{(p)}=(a_{0}+a_{1}r+\cdots+a_{p}r^{2p})r^{\delta}
\exp\left[-\frac{\sqrt{a}r^2+\sqrt{d}r^{-2}}{2}\right], \eqno(51)$$
where $a_{i} (i=0, 1, \cdots, p)$ can be evaluated from recursion relation 
Eq. (40) and the normalization condition. 

\vskip 1cm
\begin{center}
{\large 5. Concluding Remarks}
\end{center}

In this paper, applying an ansatz to the eigenfunction, we 
have obtained the exact solutions of the
two-dimensional Schr\"{o}dinger equation with the certain 
potentials such as the sextic potential $V(r)=ar^2+br^4+cr^6$, the mixed 
potential $V(r)=ar+br^2+cr^{-1}$ as well as the singular even-power potential
$V(r)=ar^2+br^{-2}+cr^{-4}+dr^{-6}$, respectively. 
The corresponding restrictions on the parameters of the 
potential and the angular momentum $m$ have been obtained for
the different potentials. 
The study for other classes of certain central potentials by this method 
is in progress.

\vskip 1cm 
{\bf Acknowledgments}. The author gratefully acknowledges Professor Zhong-Qi Ma
for encouragement and Professor Mark. S. Child for his hospitable
invitation to University of Oxford and his nice suggestions to the
manuscript. I am also grateful to 
Dr. Xiao-Gang Wang for help in Oxford. 
This work is supported 
in part by the Royal Society and in part 
by Division of Chemical Sciences, Office of Basic Energy Sciences, 
Office of Energy Research, US Department of Energy.


\end{document}